\documentstyle[preprint,aps,floats,psfig]{revtex}
\begin{document}
\newenvironment{tab}[1]
{\begin{tabular}{|#1|}\hline}
{\hline\end{tabular}}

\title {Proximity effect in superconductor Aharonov Bohm loop
hybrid structures}

\author{N. Stefanakis}
\address{Universit\"at  T\"ubingen,
Institut f\"ur Theoretische Physik,
Auf der Morgenstellen 14, 72076
T\"ubingen Germany}  
\date{\today}
\maketitle

\begin{abstract}
We study the proximity 
effect in superconductor - 
metallic or ferromagnetic Aharonov Bohm loop hybrid structures
self consistently using the Bogoliubov-deGennes formalism within
the two dimensional extended Hubbard model.
We calculate the local density of states,
the pair amplitude, and persistent currents as a function of 
several modulation parameters, the position around the loop, the 
exchange field, the magnetic flux through the loop, 
the loop size, and the chemical potential in the 
loop atoms. 
We find that the parameters above 
can modulate the 
proximity effect.
\end{abstract}
\pacs{}
\section{Introduction}

Recently the proximity effect has been probed as
decaying oscillations of the 
density of states in $s$-wave superconductor ferromagnet hybrid 
structures \cite{kontos} and a phase shift of half flux quantum in the 
diffraction pattern of a ferromagnetic $0-\pi$ SQUID \cite{guichard}.
Similar effects have been observed in $d$-wave 
\cite{freamat,freamat1} superconductor ferromagnet hybrid structures.
Theoretical explanation has been given in the framework of the 
quasiclassical theory for $s$-wave \cite{zareyan1} 
and $d$-wave case \cite{zareyan2}. 
In these structures the exchange field
modulates the period of the pair amplitude oscillations. 

Moreover much interest has been focused recently on the 
manipulation of entangled states which are formed by extracting 
Cooper pairs from the superconductor. For example a beam 
splitter has been proposed \cite{lesovik} and also several experiments 
that involve ferromagnetic electrodes connected to 
superconductors \cite{melin,jirari,stefan2}. 
These structures have acquired considerable 
interest the last years due to the possibility to use 
the $\pi$ states in solid state qubit implementation.

Moreover great experimental efforts have been devoted the last
years to control the conductance in mesoscopic structures by
phase sensitive methods. 
For example the differential
conductance of a mesoscopic metallic loop in contact with
two superconducting electrodes has been examined \cite{courtois}.

In this 
paper 
our goal is to explore several new aspects
related to the control of the proximity effect in
superconductor -
Aharonov Bohm loop (AB) structures.
The basic quantities which we calculate are
the local density of states (LDOS), the pair amplitude, 
and persistent currents, 
as a function of 
the magnetic field, exchange field, loop size, 
position in the loop, and chemical potential in the loop atoms.
We find that the quasiparticle properties of the hybrid structure 
are modulated by these parameters. 
The method that we use is based on exact diagonalizations
of the Bogoliubov-de Gennes equations associated to the mean field
solution 
of an extended Hubbard model.
Our predictions from the simulations of this model are of interest in
view of future STM spectroscopy experiments on nanostructures.

The article is organized as follows. In Sec. II we
develop the model and discuss the formalism. 
In Sec. III we discuss the effect of the magnetic field on the LDOS. 
In Sec. IV we present the results for the calculation of the 
currents. 
Finally summary and discussions are presented in the last section.

\section{BdG equations
within the Hubbard model}

The Hamiltonian for the Hubbard lattice model
is
\begin{eqnarray}
H & = & -t\sum_{<i,j>\sigma}c_{i\sigma}^{\dagger}c_{j\sigma} 
+\mu \sum_{i\sigma} n_{i\sigma}
+\sum_{i\sigma} h_{i\sigma}n_{i\sigma} \nonumber \\
  & + & V_0\sum_{i} n_{i\uparrow} n_{i\downarrow}
,~~~\label{bdgH}
\end{eqnarray}
where $i,j$ are sites indices and the angle brackets indicate that the 
hopping is only to nearest neighbors, 
$n_{i\sigma}=c_{i\sigma}^{\dagger}c_{i\sigma}$ is the electron number 
operator in site $i$, $\mu$ is the chemical potential.
$h_{i\sigma}=-h\sigma_z$, is the exchange field 
in the ferromagnetic region
and $\sigma_z=\pm 1$ is the eigenvalue of the 
$z$ component of the Pauli matrix. 
$V_0$ is
the on site interaction strength which gives rise to 
superconductivity.
Within the mean field approximation Eq. (\ref{bdgH}) is reduced  to 
the Bogoliubov deGennes equations \cite{gennes}:
\begin{equation} 
\left(
\begin{array}{ll}
  \hat{\xi} & \hat{\Delta} \\
  \hat{\Delta}^{\ast} & -\hat{\xi} 
\end{array}
\right)
\left(
\begin{array}{ll}
  u_{n \uparrow}(r_i) \\
  v_{n \downarrow}(r_i) 
\end{array}
\right)
=\epsilon_{n\gamma_1}
\left(
\begin{array}{ll}
  u_{n \uparrow}(r_i) \\
  v_{n \downarrow}(r_i) 
\end{array}
\right)
,~~~\label{bdgbdg1}
\end{equation}

\begin{equation} 
\left(
\begin{array}{ll}
  \hat{\xi} & \hat{\Delta} \\
  \hat{\Delta}^{\ast} & -\hat{\xi} 
\end{array}
\right)
\left(
\begin{array}{ll}
  u_{n \downarrow}(r_i) \\
  v_{n \uparrow}(r_i) 
\end{array}
\right)
=\epsilon_{n\gamma_2}
\left(
\begin{array}{ll}
  u_{n \downarrow}(r_i) \\
  v_{n \uparrow}(r_i) 
\end{array}
\right)
,~~~\label{bdgbdg2}
\end{equation}

such that 
\begin{equation}
\hat{\xi}u_{n\sigma}(r_i)=-t\sum_{\hat{\delta}} 
u_{n\sigma}(r_i+\hat{\delta})+(\mu^I(r_i)+\mu)u_{n\sigma}(r_i)+
h_i\sigma_z u_{n\sigma}(r_i),~~~\label{bdgxi}
\end{equation}

\begin{equation}
\hat{\Delta}u_{n\sigma}(r_i)=\Delta_0(r_i)u_{n\sigma}(r_i), 
~~~\label{bdgdelta}
\end{equation}
where the pair potential is defined by
\begin{equation}
\Delta_0(r_i)\equiv 
V_0<c_{\uparrow}(r_i)c_{\downarrow}(r_i)>.~~~\label{bdgdelta0}
\end{equation}
Equations 
(\ref{bdgbdg1},\ref{bdgbdg2}) 
are subject to the self consistency requirement 
\begin{equation}
\Delta_0(r_i)  =  V_0(r_i)F(r_i)= 
V_0(r_i)\sum_{n} \left[ 
u_{n\uparrow}(r_i)v_{n\downarrow}^{\ast}(r_i)(1-f(\beta
\epsilon_{n\gamma_1}))+ 
u_{n\downarrow}(r_i)v_{n\uparrow}^{\ast}(r_i)f(\beta
\epsilon_{n\gamma_2}) \right]
,~~~\label{bdgselfD0}
\end{equation}
$F(r_i)$ is the pair amplitude. 
We solve the above equations self consistently. 
The numerical procedure has been described elsewhere
\cite{stefan2,stefan,tanuma,zhu}.
 
The LDOS at the $i$th site is given by
\begin{equation}
\rho_i(E)=-\sum_{n\sigma} 
\left [ |u_{n\sigma}(r_i)|^2 f^{'}(E-\epsilon_n) 
+ |v_{n\sigma}(r_i)|^2 f^{'}(E+\epsilon_n) \right ]
,~~~\label{bdgdos}
\end{equation}
where $f^{'}$ is the derivative of the Fermi function,
\begin{equation}
f(\epsilon)=\frac{1}{\exp(\epsilon/k_B T) + 1}
\end{equation}.

\section{LDOS as a function of the magnetic field}

\begin{figure}
\begin{center}
\leavevmode
\psfig{figure=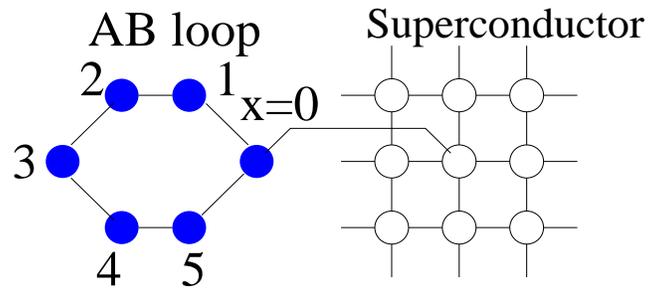,width=8.5cm,angle=0}
\end{center}
\caption{The junction of the normal metal or ferromagnetic AB loop with 
the superconductor. The numbering of the sites along 
the loop is illustrated.}
\label{ABloop.fig}
\end{figure}

We demonstrate in this section that the magnetic flux through a
metallic or ferromagnetic AB loop
which is connected to a superconductor
as seen in Fig. \ref{ABloop.fig}
can be used to control the proximity effect in 
this hybrid structure.  
The magnetic flux through the loop is modeled as a factor 
$e^{i\phi}$ where $\phi=2\pi \Phi/ \Phi_0=2\pi f$,
in the hopping element, where $\Phi_0$ is the unit 
of the flux quantum. 
In the calculation we used a small cluster 
of $7\times7$ sites to model the superconductor and 
$6$ sites to model the metallic or ferromagnetic loop.
The AB effect appears as periodic oscillations of the conductance 
of a ring as a function of the enclosed magnetic flux $f$. 

\subsection{isolated AB loop}

\begin{figure}
\begin{center}
\leavevmode
\psfig{figure=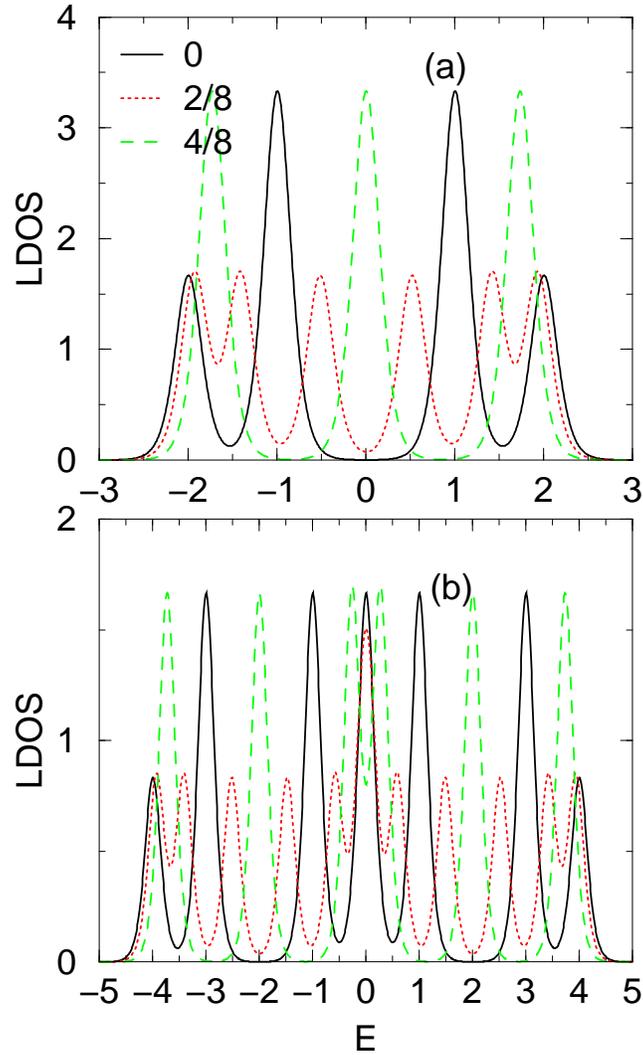,width=8.5cm,angle=0}
\end{center}
\caption{
(a) The LDOS at $E=0$ 
as a function of the energy, for different 
values of the magnetic flux $f=0,2/8,4/8$, 
for site $0$ of the AB loop, for $h=0$. The pair 
interaction is $V_0=0$ and the hopping between the AB ring and the 
two dimensional reservoir is zero. 
(b) The same as in (a) but for $h=2$.}
\label{abldosft0.fig}
\end{figure}

We first study the case of an 
isolated AB loop. Magnetoresistance oscillations with period 
$h/e$ have been observed in Au ring along with weaker 
$h/2e$ oscillations \cite{webb}. Also magnetoresistance oscillations of 
period $h/2e$ (corresponding to a superconducting flux quantum) 
were observed on a two dimensional honeycomb Mg network of loops 
and were attributed to weak localization effects \cite{pannetier}. 
We present in Fig. \ref{abldosft0.fig} the LDOS versus energy
for site $0$, for different values of the flux. The pair
interaction in the superconductor is zero, and the 
ring is disconnected from the normal metal reservoir 
namely the corresponding 
hopping element of the tight binding Hamiltonian is zero. 
We see that the LDOS shows oscillations as a function of the 
applied magnetic flux $f$. 
Also the LDOS for finite exchange field can be obtained 
by adding the LDOS for spin up and spin down Hamiltonian, 
so the number of peaks is double. 

\subsection{AB loop in contact with normal metal reservoir}

\begin{figure}
\begin{center}
\leavevmode
\psfig{figure=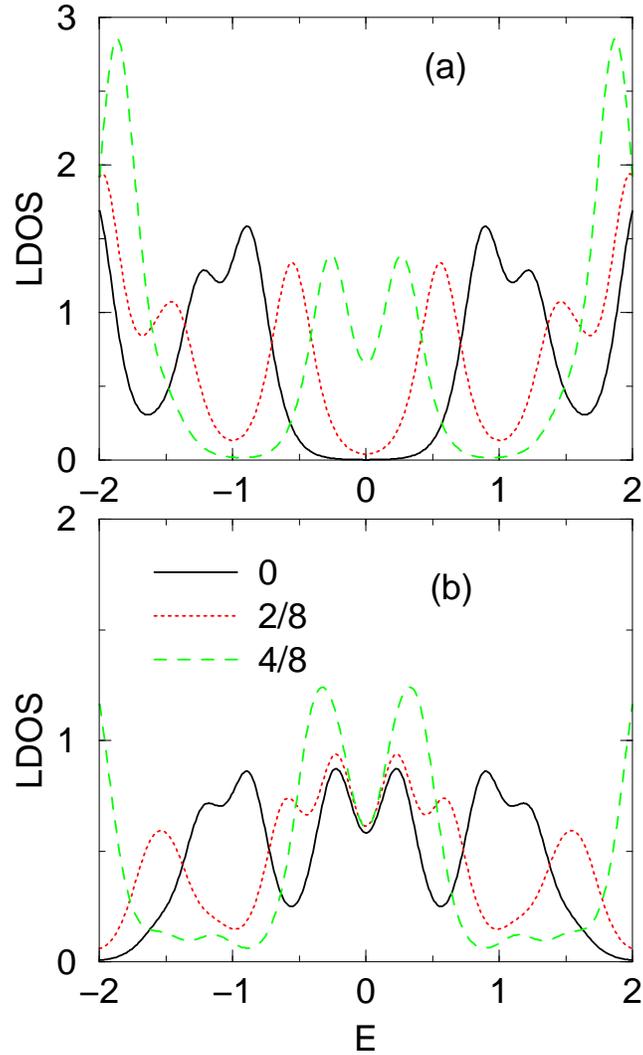,width=8.5cm,angle=0}
\end{center}
\caption{
(a) The LDOS at $E=0$ 
as a function of the energy, for different 
values of the magnetic flux $f=0,2/8,4/8$, 
for site $0$ of the AB loop, for $h=0$. The pair 
interaction is $V_0=0$, but the coupling between the 
ring and the normal metal reservoir is finite. 
(b) The same as in (a) but for $h=2$.}
\label{abldosfV0.fig}
\end{figure}

Secondly we would like to make a comparison of the results derived 
above with the case where AB loop is connected to a 
normal metal reservoir. 
We present in Fig. \ref{abldosfV0.fig} the LDOS versus energy 
for site $0$, for different values of the flux.
The basic results are summarized as 
follows. The LDOS shows again modulation with the applied flux. 
However the form of the LDOS is changed due to the exchange of 
electrons between the ring and the reservoir.  
For all values of the enclosed flux it presents a $U$-like form 
which is characteristic for chain of atoms. 

\subsection{superconductor-metallic AB loop}

\begin{figure}
\begin{center}
\leavevmode
\psfig{figure=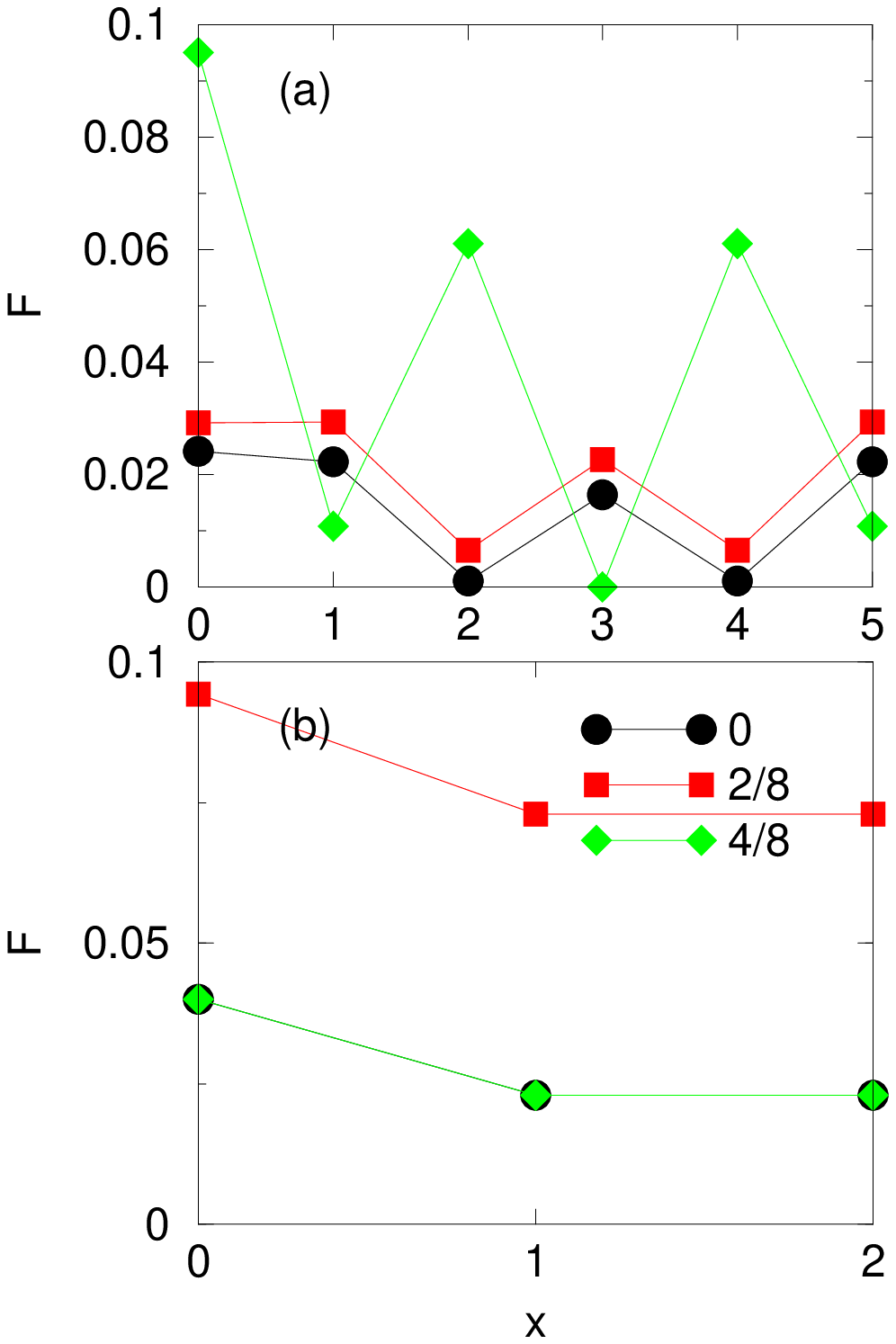,width=8.5cm,angle=0}
\end{center}
\caption{
(a) The pairing amplitude along the AB loop of $6$ sites,
which is in contact with 
a superconducting reservoir,
for different values of the magnetic flux in the loop 
$f=0,2/8,4/8$, 
and exchange field equal to $h=0$.
(b) The same as in (a) but for a $3$ site AB loop.}
\label{abpah0.fig}
\end{figure}

\begin{figure}
\begin{center}
\leavevmode
\psfig{figure=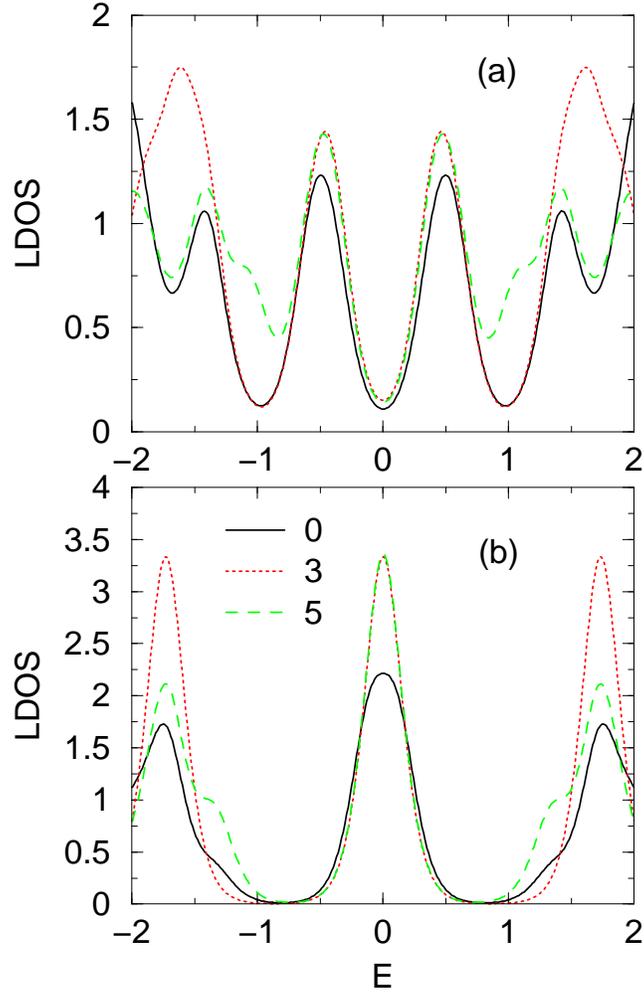,width=8.5cm,angle=0}
\end{center}
\caption{
(a) The LDOS
as a function of energy,
for different sites along the AB loop $x=0,3,5$
for magnetic flux equal to $f=2/8$, and exchange field 
equal to $h=0$.
(b) The same as in (a) but for $f=4/8$.}
\label{abldosxh0.fig}
\end{figure}

\begin{figure}
\begin{center}
\leavevmode
\psfig{figure=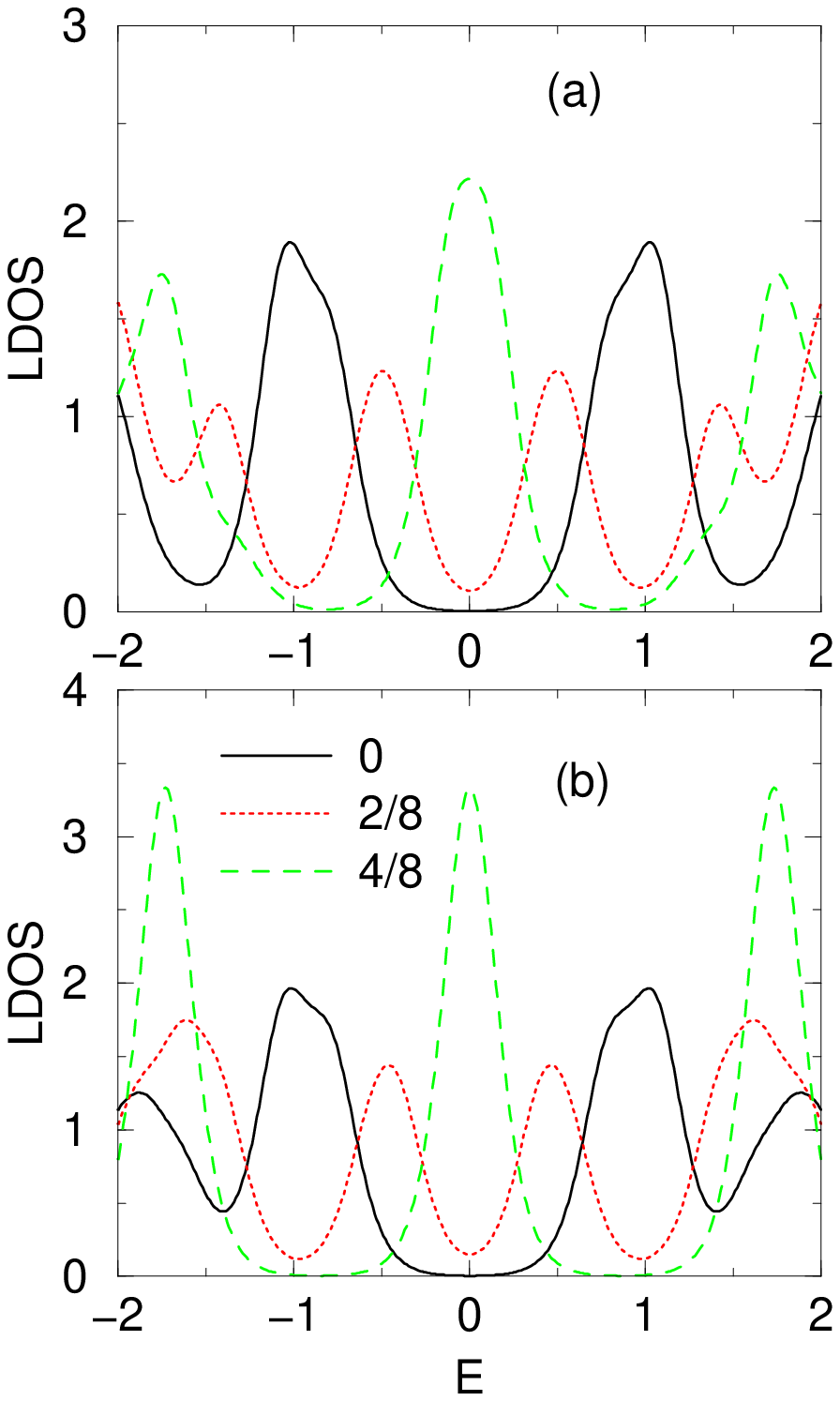,width=8.5cm,angle=0}
\end{center}
\caption{
(a) The LDOS
as a function of energy,
for site $0$ of the AB loop, for different values of the 
magnetic flux $f=0,2/8,4/8$,
and exchange field 
equal to $h=0$.
(b) The same as in (a) but for the site $3$.}
\label{abldosphih0.fig}
\end{figure}

We discuss in the subsection the case where a metallic AB loop is connected 
to a superconductor. The quasiparticle properties in the ring 
are modified by the proximity effect. 
In the usual proximity effect in the superconductor
normal metal bilayer the pair amplitude
decays away from the
interface inside the normal metal.
However in the case of a normal metal loop in contact 
with a superconductor, 
the pair amplitude decays inside the loop, 
symmetrically around the site where the superconductor 
is connected to the loop
(see the symmetry between sites $1$ and $5$ 
Fig. \ref{abpah0.fig}(a) for a $6$ sites loop 
and between sites $1,2$ in 
Fig. \ref{abpah0.fig}(b), for a $3$ sites loop). 
The pair amplitude changes with the magnetic flux.
This may be related to size effects.
The evolution of the LDOS around the 
loop is not periodic. 
The LDOS shows the $U$-like form which is characteristic 
of a line of atoms which changes 
as we move around the loop as seen in Fig. \ref{abldosxh0.fig}(a).
Especially for $f=4/8$ it shows 
a pic at zero energy
(see Fig. \ref{abldosxh0.fig}(b)). This means that the proximity effect 
becomes long ranged one for this particular value of the magnetic flux.  
However the evolution with respect to the 
enclosed magnetic flux is periodic.
This is seen in Fig. \ref{abldosphih0.fig} where 
the LDOS is presented for fixed site on the loop but for 
different values of the magnetic flux. We present only 
$f<4/8$ due to symmetry. 
So the proximity effect can be modulated by the AB magnetic flux.

\subsection{superconductor-ferromagnetic AB loop} 
\begin{figure}
\begin{center}
\leavevmode
\psfig{figure=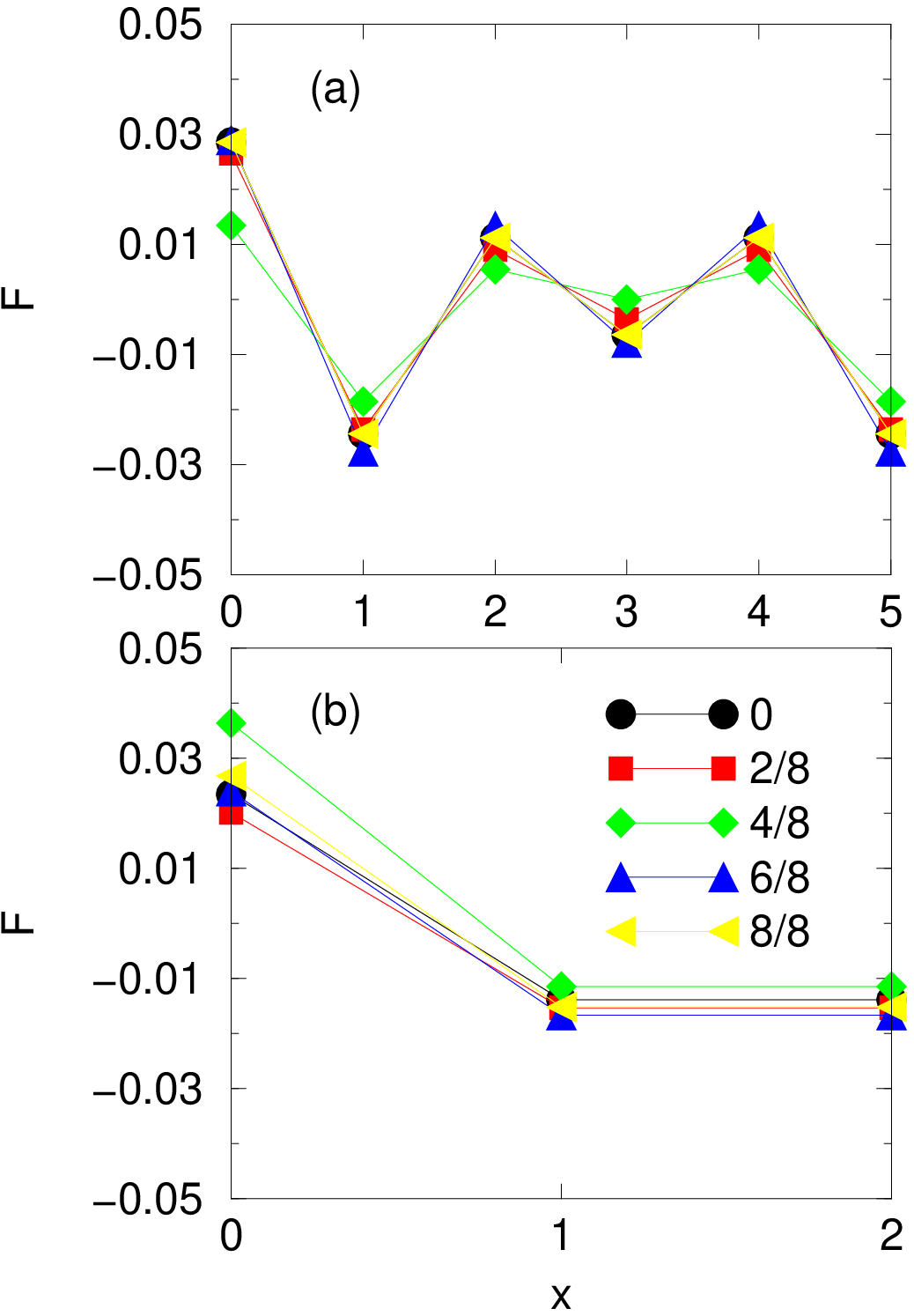,width=8.5cm,angle=0}
\end{center}
\caption{
(a) The pairing amplitude along the AB loop of 
$6$ sites, 
for different values of the magnetic flux in the loop 
$f=0,2/8,4/8,6/8,8/8$, 
and exchange field equal to $h=2$.
(b) The same as in (a) but for a $3$ site AB loop.}
\label{abpa.fig}
\end{figure}

We now study a ferromagnetic AB loop in contact with 
a superconductor.
When the magnetic flux is zero the proximity effect 
appears in the loop sites as oscillations of the pairing 
amplitude inside the ferromagnetic material. 
In the usual proximity effect in the superconductor 
ferromagnet bilayer the proximity effect appears as 
decaying oscillations of the pair amplitude with 
alternating sign away from the 
interface inside the ferromagnetic material. 
However 
due to the annular geometry of our structure the proximity effect 
appears symmetrically around the loop (note the symmetry 
between sites $1$ and $5$ in Fig. \ref{abpa.fig}(a) for the 
$6$ site loop and between sites $1$ and $2$ for the $3$ site
loop structure in Fig. \ref{abpa.fig}(b)). A second 
difference is that the proximity effect oscillations 
for very small loop sizes are not decaying but are rather 
symmetric. 
As indicated in these figures the oscillations of pair 
amplitude inside the ferromagnetic material change 
as we change magnetic flux that is applied through the loop.
The pair amplitude function returns to itself after one 
period in $\Phi_0$ is completed. The flux can change the 
amplitude of the oscillations but not the period. 

\begin{figure}
\begin{center}
\leavevmode
\psfig{figure=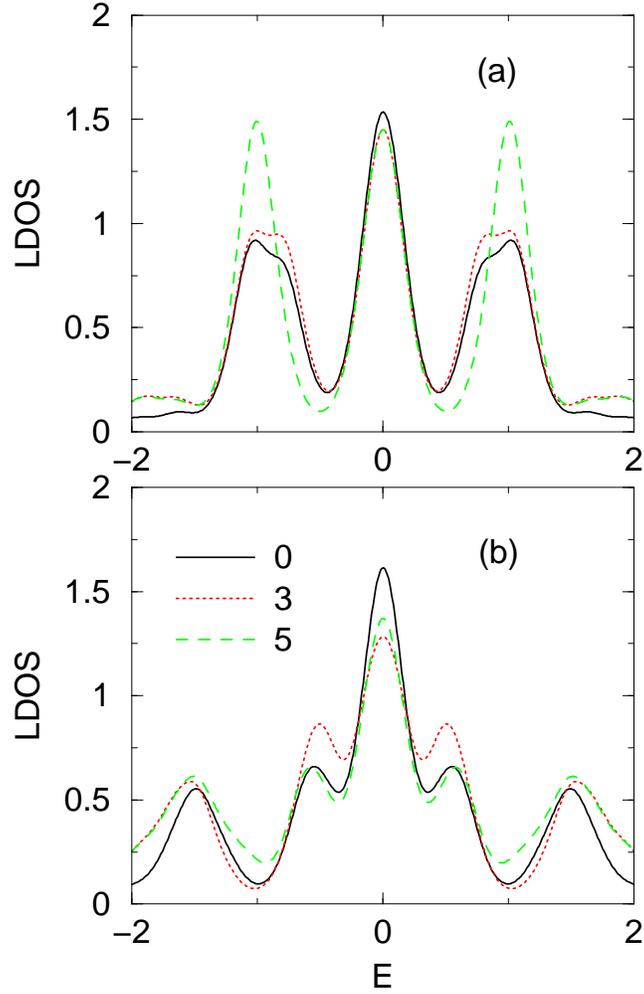,width=8.5cm,angle=0}
\end{center}
\caption{
(a) The LDOS
as a function of energy, for a two dimensional electron gas,
for different sites along the AB loop $x=0,3,5$
for magnetic flux equal to $f=0$, and exchange field 
equal to $h=2$.
(b) The same as in (a) but for $f=2/8$.}
\label{abldosx.fig}
\end{figure}

\begin{figure}
\begin{center}
\leavevmode
\psfig{figure=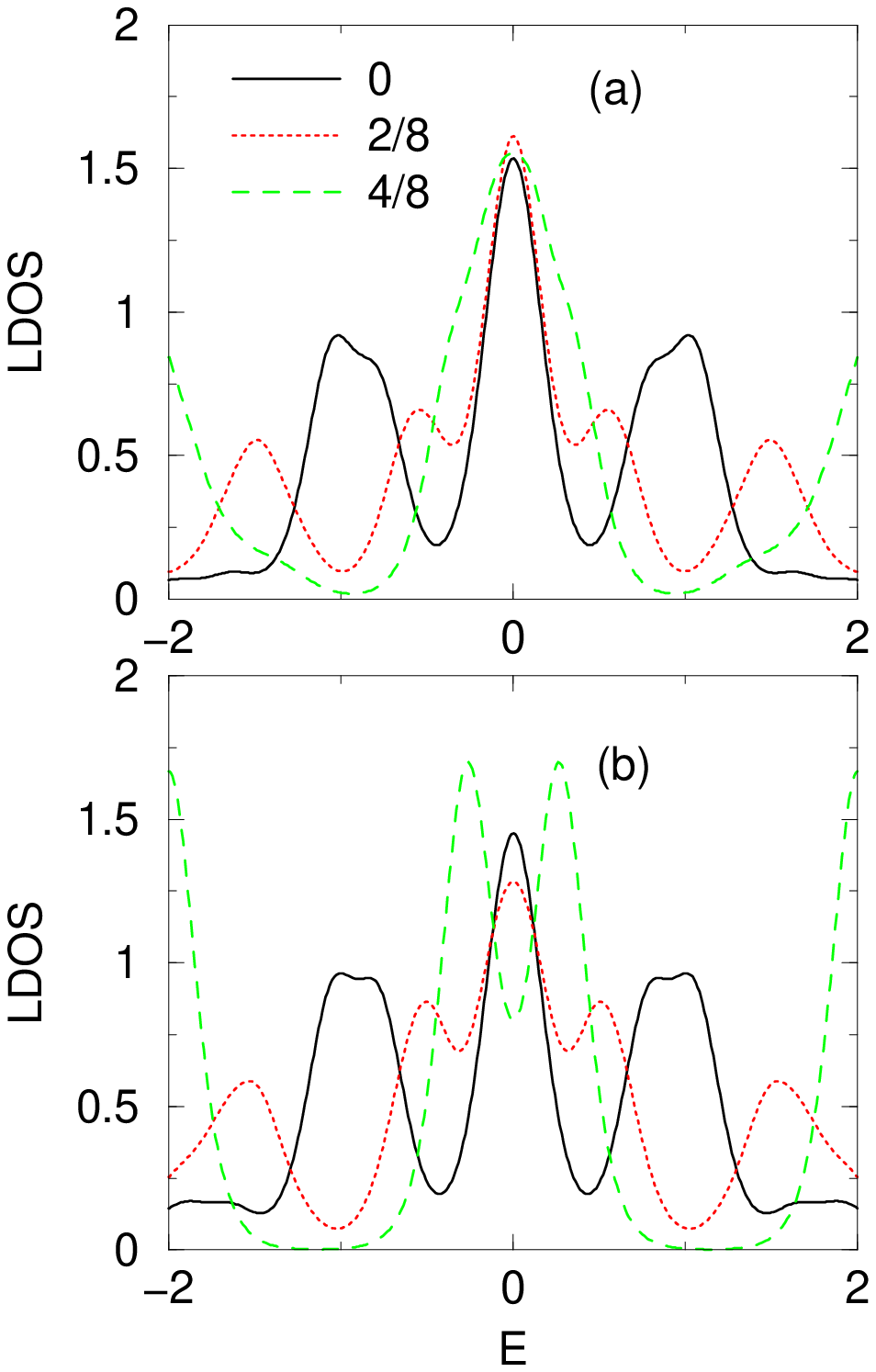,width=8.5cm,angle=0}
\end{center}
\caption{
(a) The LDOS
as a function of energy,
for site $0$ of the AB loop, for different values of the 
magnetic flux $f=0,2/8,4/8$,
and exchange field 
equal to $h=2$.
(b) The same as in (a) but for the site $3$.}
\label{abldosphi.fig}
\end{figure}

\begin{figure}
\begin{center}
\leavevmode
\psfig{figure=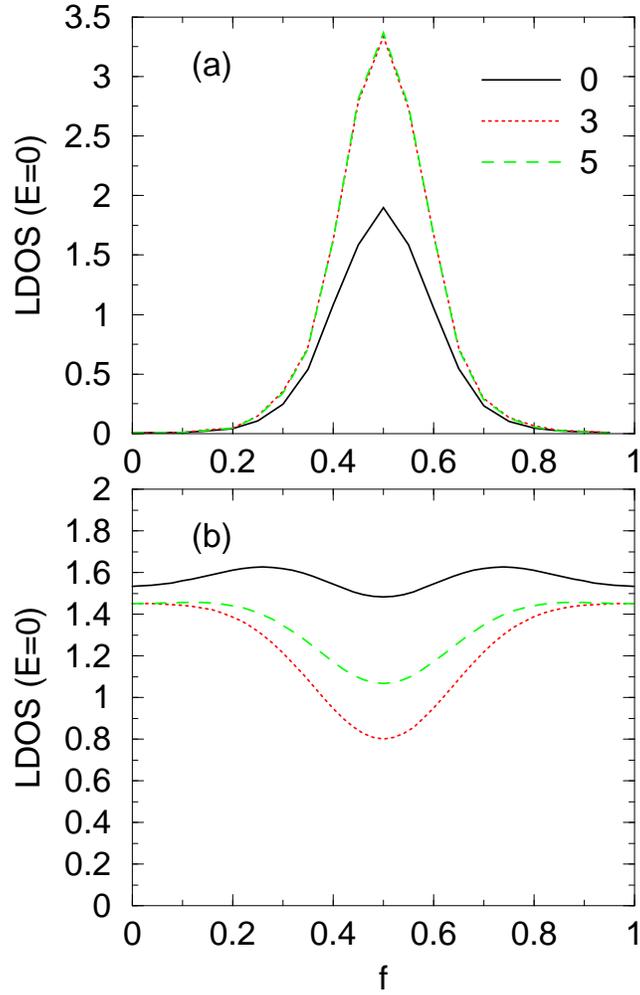,width=8.5cm,angle=0}
\end{center}
\caption{
(a) The LDOS at $E=0$ 
as a function of the magnetic field,
for sites $0,3,5$ of the AB loop, for $h=0$.
(b) The same as in (a) but for $h=2$.}
\label{abldosf.fig}
\end{figure}

The modulation of the oscillations of the pair amplitude 
with $f$ induces additional effects in the LDOS seen in 
Fig. \ref{abldosx.fig}. When $f=0$
the ZEP in the LDOS practically
does not change around the loop as seen in 
Fig. \ref{abldosx.fig}(a). This means that the proximity effect 
becomes long ranged as in the metallic loop case. 
The presence of the pic at $f=0$ which is absent in the 
case where $h=0$ is attributed to the exchange field. 
Namely the effect of the 
ferromagnetism is to shift the LDOS that corresponds to $h=0$ by
an energy $E=h$, so that a peak appears at zero energy when the 
van Hove singularity crosses the Fermi energy.  
Also when $f=2/8$ 
(see Fig. \ref{abldosx.fig}(b)) 
a peak develops at zero energy.
Moreover the LDOS is a periodic function 
of the applied magnetic flux as seen in Fig. \ref{abldosphi.fig}
and
it shows a different form for $h=0$ and $h=2$ (see Fig. \ref{abldosf.fig}).

Comparing with 
Figs. \ref{abldosft0.fig} and \ref{abldosfV0.fig} we see that 
the main effect of the superconductor is to enhance the LDOS 
and also to induce long range proximity effects for certain 
values of $f$, e.g. $f=4/8$. The long range proximity effect 
appears in the LDOS as a peak at zero energy for some values 
of $f$. 

\section{persistent currents}

We would like in this section to describe the effect of the 
proximity effect on the 
persistent currents of a  small loop that encloses magnetic flux 
and is coupled to a superconductor. 
For small isolated one dimensional metallic rings 
enclosing magnetic flux, the existence of persistent currents 
has been predicted in theory \cite{cheung} 
and verified in magnetization response of isolated Cooper 
rings to a slowly varying magnetic flux \cite{levy}. 
In isolated loops the relation of the persistent current as a function 
of the flux $I(f)$ is modulated by the loop circumference and the 
chemical potential. The relation $I(f)$ is periodic with period $\Phi_0$ and in 
some cases $\Phi_0/2$. 

We study first the case of an isolated AB loop. 
In this case one can give simple physical argument for the 
persistent currents as follows. The 
phase coherence of the wave function enters the calculation through the 
flux modulated boundary conditions
\begin{equation}
\Psi_n(x+L)=\exp [2i\pi f]\Psi_n(x),
\end{equation}
where $L$ is the loop circumference.
Then the current is 
$I_n=\frac{e v_n}{L}$ where 
$v_n=\frac{1}{\hbar} \frac{\partial E_n}{\partial k_n}$.
Then by identifying $2\pi f$ and $kL$ we have 
$I_n=\frac{e}{h}\frac{\partial E_n}{\partial f}$. 

\begin{figure}
\begin{center}
\leavevmode
\psfig{figure=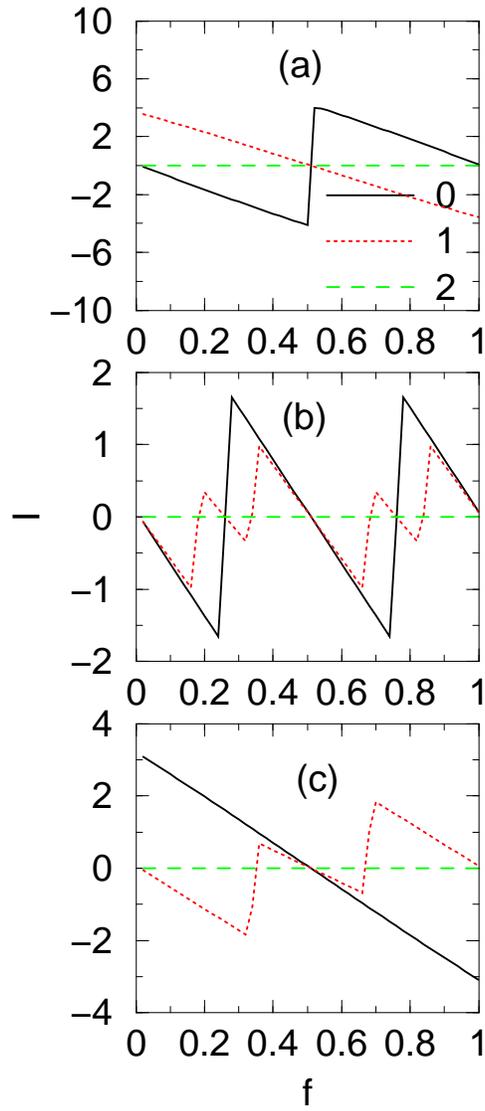,width=6.5cm,angle=0}
\end{center}
\caption{
The current - flux relation for different values of the 
exchange field $h=0,1,2$ and different number of sites in the 
ring (a) 6, (b) 7, (c) 8. The pair interaction is $V_0=0$ 
and the hopping between the AB ring and the
two dimensional reservoir is zero.}
\label{ringvaryl.fig}
\end{figure}

In our case the energy spectrum is obtained from the 
eigenvalues of the BdG Hamiltonian. 
Then the current for the $n$ state is 
$I_n=\frac{e}{h}\frac{\partial E_n}{\partial f}$.
To calculate the total current we perform a summation over 
the eigenvalues of the system. 
We present in Fig. \ref{ringvaryl.fig} the $I(f)$ relation for
$V_0=0$ and hopping elements connecting the AB loop and 
the normal metal reservoir equal to zero, 
for different loop sizes and exchange fields.
We see that the $I(f)$ relation changes with the loop size 
and exchange field
depending on the number of electrons in the loop and 
presents linear variation with the applied flux. 

\begin{figure}
\begin{center}
\leavevmode
\psfig{figure=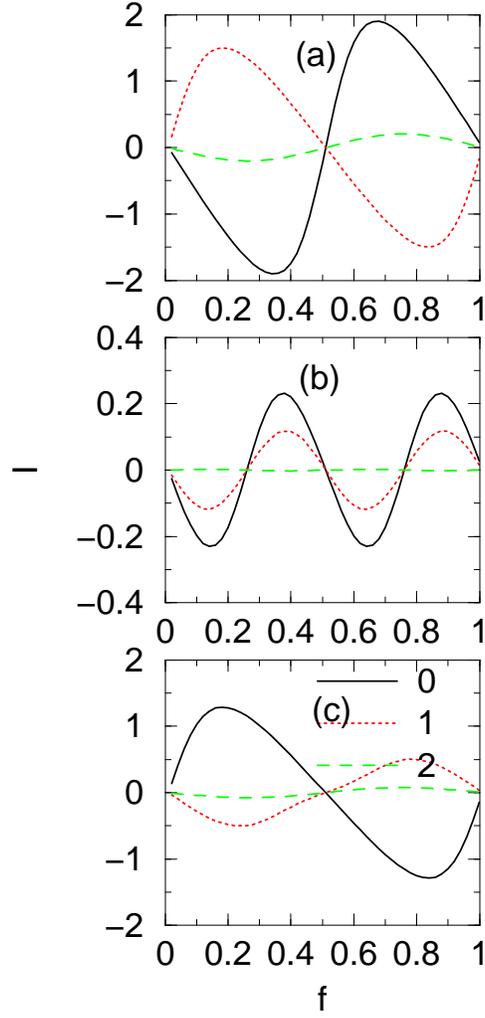,width=6.5cm,angle=0}
\end{center}
\caption{
The current - flux relation for different values of the 
exchange field $h=0,1,2$ and different number of sites in the 
ring (a) 6, (b) 7, (c) 8. The pair interaction is $V_0=0$ 
and the ring is coupled to a normal metal reservoir.}
\label{supra0varyl.fig}
\end{figure}

We
study then the case where $V_0$ equals zero,
and the AB loop is connected to a normal metal reservoir. 
We present in Fig. \ref{supra0varyl.fig} the $I(f)$ relation
for different loop sizes and exchange fields.
We see that the $I(f)$ relation changes with the loop size. 
Also for fixed loop size the $I(f)$ relation changes sign 
as we increase the exchange field and the critical current 
is reduced for large 
values of the exchange field. Moreover due to the presence 
of the electron exchange with the reservoir the $I(f)$ relation 
is not linear but is rounded. 

\begin{figure}
\begin{center}
\leavevmode
\psfig{figure=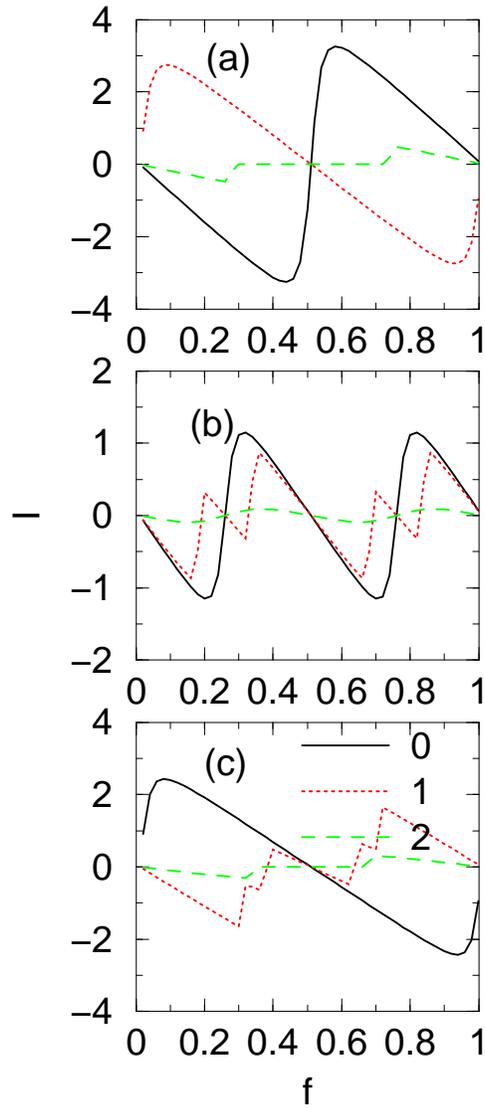,width=6.5cm,angle=0}
\end{center}
\caption{
The current - flux relation for different values of the 
exchange field $h=0,1,2$ and different number of sites in the 
ring (a) 6, (b) 7, (c) 8. The pair interaction in the reservoir is $V_0=-3.5$}
\label{varyl.fig}
\end{figure}

We now turn to the case where the proximity effect is present. 
The pairing interaction in the superconductor is set equal to 
$V_0=-3.5$. We present in Fig. \ref{varyl.fig} the $I(f)$ relation for
different loop sizes and exchange fields.
We see that the $I(f)$ relation is enhanced comparable to the 
case where $V_0=0$. Also the period of the $I(f)$ relation 
changes with the exchange field. 

\begin{figure}
\begin{center}
\leavevmode
\psfig{figure=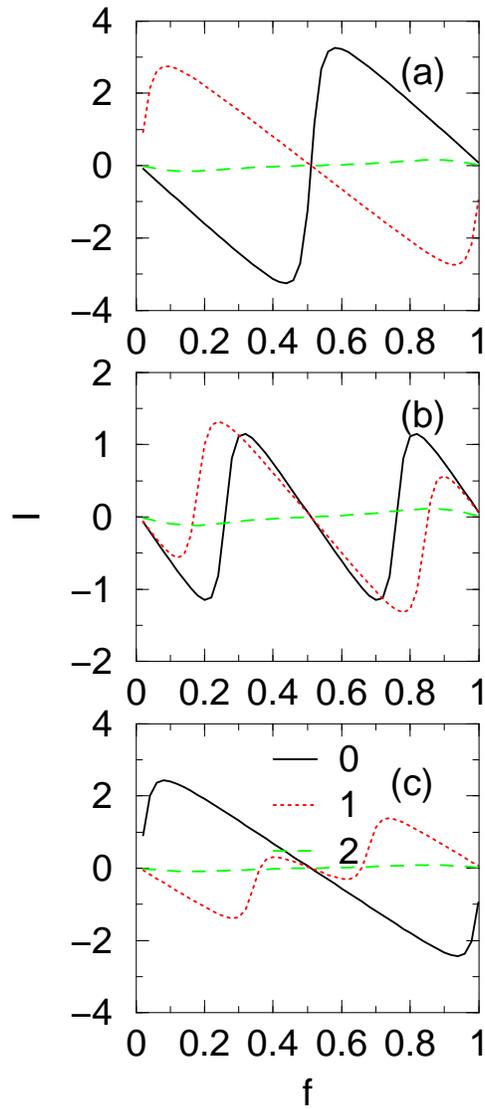,width=6.5cm,angle=0}
\end{center}
\caption{
The current - flux relation for different values of the 
chemical potential $\mu=0,1,2$ in the loop atoms
and different number of sites in the 
ring (a) 6, (b) 7, (c) 8. The exchange field is $h=0$, 
and the pair interaction in the reservoir is $V_0=-3.5$.}
\label{varymu.fig}
\end{figure}

The modulation of the $I(f)$ relation with the exchange field is 
explained by the change of the number of electrons in the 
ring with the exchange field. Similar effects we observe when 
the chemical potential for the ring atoms is considered 
as modulation parameter. We present in 
Fig. \ref{varymu.fig} the $I(f)$ relation for
$V_0=-3.5$ and no exchange field 
for different loop sizes for different values of the 
chemical potential in the loop atoms. We 
see again that the $I(f)$ relation changes 
with the chemical potential in the loop atoms.

\section{conclusions}
We calculated the LDOS, the pair amplitude, and 
the persistent currents for
superconductor - AB loop hybrid structures, 
within the extended lattice Hubbard model.  
The quasiparticle properties depend on the 
loop size the magnetic flux, the exchange field and 
the chemical potential in the loop atoms. 
The proximity effect modifies the LDOS and for particular values of the 
magnetic flux it shows long range character. 
Also it enhances the persistent currents 
in the loop and modifies in some cases the $I(f)$ relation.

\bibliographystyle{prsty}

\begin{thebibliography}{99}

\bibitem{kontos} T. Kontos M. Aprili, J. Lesueur, and X. Grison,
Phys. Rev. Lett. {\bf 86}, 304 (2001).

\bibitem{guichard} W. Guichard, M. Aprili, O. Bourgeois, T. Kontos, 
J. Lesueur, and P. Gandit,
Phys. Rev. Lett. {\bf 90}, 167001 (2003).

\bibitem{freamat} M. Freamat and K.-W. Ng,
cond-mat/0301081.

\bibitem{freamat1} M. Freamat and K.-W. Ng,
cond-mat/0305446.

\bibitem{zareyan1} M. Zareyan, W. Belzig, and Yu.V. Nazarov,
Phys. Rev. Lett. {\bf 86}, 308 (2001).

\bibitem{zareyan2} Z. Faraii and M. Zareyan,
cond-mat/0304336.

\bibitem{lesovik} G.B. Lesovik, T. Martin, and G. Blatter,
Eur. Phys. J. B {\bf 24}, 287 (2001).

\bibitem{melin} R. M\'elin and D. Feinberg,
Eur. Phys. J. B {\bf 26}, 101 (2002).

\bibitem{jirari} H. Jirari, R. M\'elin and N. Stefanakis,
Eur. Phys. J. B {\bf 31}, 125 (2003).

\bibitem{stefan2} N. Stefanakis and R. M\'elin,
J. Phys. Condens. Matter {\bf 15}, 3401 (2003).

\bibitem{courtois} H. Courtois, Ph. Gandit, D. Mailly, and B. Pannetier,
Phys. Rev. Lett. {\bf 76}, 130 (1996).

\bibitem{gennes} P.G. de Gennes, {\em Superconductivity of Metals and Alloys} 
(Benjamin, New York, 1966).

\bibitem{stefan} N. Stefanakis, Phys. Rev. B
{\bf 66}, 024514 (2002).

\bibitem{tanuma} Y. Tanuma, Y. Tanaka, M. Yamashiro, and S. Kashiwaya,
Phys. Rev. B
{\bf 57}, 7997 (1998).

\bibitem{zhu} J.-X. Zhu and C.S. Ting,
Phys. Rev. B
{\bf 61}, 1456 (2000).

\bibitem{webb} R.A. Webb, S. Washburn, C.P. Umbach, and R.B. Laibowitz,
Phys. Rev. Lett. {\bf 54}, 2696 (1985).

\bibitem{pannetier} B. Pannetier, J. Chaussy, R. Rammal, and P. Gandit,
Phys. Rev. Lett. {\bf 53}, 718 (1984).

\bibitem{cheung} H.-F. Cheung, Y. Gefen, E.K. Riedel, and W.-H. Shih,
Phys. Rev. B
{\bf 37}, 6050 (1988).

\bibitem{levy} L.P. L\'evy, G. Dolan, J. Dunsmuir, and H. Bouchiat,
Phys. Rev. Lett.
{\bf 64}, 2074 (1990).


\end{thebibliography}

\end{document}